\documentclass[journal]{IEEEtran}
\usepackage{amsthm,amssymb,amsbsy,amsmath,dsfont,latexsym,dsfont,array,layout,graphicx,mathrsfs,txfonts,amsfonts,bm,subfig,color}
\usepackage{algorithm, algorithmic}
\usepackage{diagbox, tikz, pgfplots} \usetikzlibrary{plotmarks}
\usepackage{threeparttable}
\usepackage{color}
\usepackage{enumitem}
\usepackage{diagbox}
\usepackage{etaremune}
\usepackage{tikz}
\usepackage{quantikz}
\usepackage[letterpaper,top=2cm,bottom=2cm,left=3cm,right=3cm,marginparwidth=1.75cm]{geometry}
\usepackage{adjustbox}
\usepackage[switch, columnwise]{lineno}
\usepackage{subfig}
\usepackage{subfloat}
\newtheorem{definition}{Definition}

\newtheorem{theorem}[definition]{Theorem}

\newtheorem{example}[definition]{Example}

\def\BibTeX{{\rm B\kern-.05em{\sc i\kern-.025em b}\kern-.08em
    T\kern-.1667em\lower.7ex\hbox{E}\kern-.125emX}}
\usepackage{balance}
\begin{document}
\title{
Towards Minimal Fault-tolerant Error-Correction Sequence with Quantum Hamming Codes
}
\author{Sha Shi, Xiao-Yang Xu, Min-Quan Cheng, Dong-Sheng Wang and Yun-Jiang Wang
\thanks{This work was supported by the National Natural Science Foundation of China (62471368, 12047503, and 12105343), the Natural Science Foundation of Guangdong Province (2023A1515010671), the Key Research and Development Project of Shannxi Province (2023-YBGY-206 and 2024GX-YBXM-069), and the Shaanxi Provincial Young Innovative Teams in Higher Education, China. (Corresponding author: Yun-Jiang Wang.)

SS, X-Y X and Y-J W, are with the School of Telecommunications Engineering, Xidian University, Xi'an, Shannxi 710071, China (email: yunjiang.w@gmail.com; xyxu510@stu.xidian.edu.cn; sshi@xidian.edu.cn).
Y-J W is also with Guangzhou Institute of Technology, Xidian University, Guangzhou 510555 and Hangzhou Institute of Technology, Xidian University, Hangzhou 311231, China.

M-Q C is with the Key Laboratory of Education Blockchain and Intelligent Technology, Ministry of Education, and the Guangxi Key Laboratory of Multi-Source Information Mining $\&$ Security, Guangxi Normal University, Guilin 541004, China (e-mail: chengqinshi@hotmail.com).

D-S W is with the Institute of Theoretical Physics, Chinese Academy of Sciences, Beijing 100190, China. (wds@itp.ac.cn)}}

\markboth{Journal of \LaTeX\ Class Files,~Vol.~18, No.~9, June~2024}%
{How to Use the IEEEtran \LaTeX \ Templates}

\maketitle

\begin{abstract}
The high overhead of fault-tolerant measurement sequences (FTMSs) poses a major challenge for implementing quantum stabilizer codes. Here, we address this problem by constructing efficient FTMSs for the class of quantum Hamming codes $[\![2^r-1, 2^r-1-2r, 3]\!]$ with $r=3k+1$ ($k \in \mathbb{Z}^+$). Our key result demonstrates that the sequence length can be reduced to exactly $2r+1$—only one additional measurement beyond the original non-fault-tolerant sequence, establishing a tight lower bound. The proposed method leverages cyclic matrix transformations to systematically combine rows of the initial stabilizer matrix and preserving a self-dual CSS-like symmetry analogous to that of the original quantum Hamming codes. This induced symmetry enables hardware-efficient circuit reuse: the measurement circuits for the first $r$ stabilizers are transformed into circuits for the remaining $r$ stabilizers simply by toggling boundary Hadamard gates, eliminating redundant hardware. For distance-3 fault-tolerant error correction, our approach simultaneously reduces the time overhead via shorting the FTMS length and the hardware overhead through symmetry-enabled circuit multiplexing. These results provide an important advance towards the important open problem regarding the design of minimal FTMSs for quantum Hamming codes and may shed light on similar challenges in other quantum stabilizer codes.
\end{abstract}


\section{Introduction}
\IEEEPARstart{Q}{uantum} computing holds the promise of solving certain problems exponentially faster than classical computers~\cite{Bravyi22}. However, quantum systems are highly fragile and prone to environmental interference, causing rapid loss of quantum information through decoherence. Realizing this potential critically depends on advances in quantum error correction (QEC), a central focus in quantum computing research~\cite{Geo20}.

Quantum error-correcting codes (QECCs) encode logical qubits into a larger physical system, drawing inspiration from classical error correction. They serve as the foundation of QEC, addressing the inherent fragility of quantum information~\cite{Shor961}. Stabilizer codes, owing to their elegant and concise representation, have emerged as the most important class of QECCs~\cite{Gottesman97}. However, the implementation of quantum error correction is itself severely impacted by noise. For example, the ancillary quantum states prepared for stabilizer measurements may themselves be unreliable. Errors in stabilizer measurement circuits, such as those in two-qubit gates, can easily propagate. Additionally, measurement devices are prone to errors, potentially causing catastrophic failures in syndrome-based quantum decoding.

Shor's seminal work established fundamental methods for fault-tolerant (FT) QEC, addressing gate faults through stabilizer measurements using cat states and transversal gates, while addressing ancilla preparation and measurement errors via repeated syndrome extraction~\cite{Shor962}. In contrast to this sequential measurement approach, Steane introduced a parallel stabilizer measurement scheme utilizing encoded ancillary states, albeit at the cost of increased qubit overhead due to the requirement of additional ancilla qubits matching the code size~\cite{Ste03}.
Recent advancements in flag-based syndrome extraction have significantly reduced this overhead. This approach requires only a few qubits per generator to track any fault that could propagate to a data error of weight greater than one, thereby enhancing the efficiency of FTQEC~\cite{Cao20,Huang21,Tan23,PP23,Ma24}.



The identification of syndrome bits as information bits allows for the protection of stabilizer measurements against measurement device faults through the application of classical coding techniques~\cite{Y14,TB14}. This approach utilizes an overcomplete generator set in conjunction with classical linear codes to construct an extended error space for syndrome bits~\cite{L25}. In this framework, Shor's strategy can be viewed as a special case where syndrome protection is achieved through the use of repetition codes. For certain highly structured quantum codes, single-shot FTQEC is achievable, enabling stabilizer measurements to be performed only once~\cite{Bom15}.


Compared to the aforementioned fault types, less attention is devoted to effective approaches for addressing internal faults on data-qubits~\cite{N20,N22}.  These internal faults, which may not spread the resulted errors directly, can evade detection by most existing measurement circuits during syndrome extraction. 
For Shor-style FTQEC, where the stabilizers are measured one at a time in sequence, internal faults are often confused with input errors, leading to diagnostic ambiguity in non-FT architectures~\cite{N20}. Consequently, the internal faults may be amplified through error correction cycles, ultimately causing error proliferation rather than suppression.
The most commonly employed method to address this type of internal faults is the repeated measurement strategy. However, this approach incurs significant time overhead.

Recently, Delfosse et al. have systematically investigated internal faults within the framework of Shor-style FTQEC in the context of weak FT error correction~\cite{N20,N22}. They demonstrated that optimized stabilizer measurement sequences of shorter length can substantially reduce the time overhead, particularly for low-distance codes. Specifically, they established distance-3 FTQEC implementations using Quantum Hamming codes $[\![7, 1, 3]\!]$\footnote{A stabilizer quantum code that encodes $k$ logical qubits into an $n$-qubit systme, with a minimum distance of $d$, is denoted as $[\![n, k, d]\!]_q$. Here, $q$ represents the dimension of the quantum system, and it is typically omitted when $q=2$.} and $[\![15, 7, 3]\!]$, achieved through Pauli measurement sequences of length 7 and 9, respectively~\cite{N20}.

The construction of optimized measurement sequences constitutes a fundamental challenge in quantum error correction (QEC), with profound implications for both theory and practice~\cite{N22}. Previous work~\cite{N20} has shown that Pauli measurement sequences can be significantly shortened by incorporating combined $X$, $Y$, and $Z$ operator measurements. However, even for the canonical class of distance-3 quantum codes—the quantum Hamming codes $[\![2^{r}-1, 2^{r}-1-2r, 3]\!]$—although exhaustive enumeration confirms that the minimal FT sequence length is $2r+1$ (including the two examples in~\cite{N20}), a systematic construction of such sequences remains elusive. Consequently, identifying minimal-length Pauli measurement sequences for distance-3 FTQEC with quantum Hamming codes persists as an open problem~\cite{N20}, and the underlying principles governing their construction are not yet fully understood.

Here, we propose an optimized Pauli measurement sequence design for distance-3 FTQEC within the weak fault-tolerance framework (Definition 1 in~\cite{N20}), focusing on Shor-style error correction for quantum Hamming codes with parameters $[\![2^r-1, 2^r-1-2r, 3]\!]$ where $r=3k+1$ ($k \in \mathbb{Z}^+$). Our key contributions are as follows.

First, \emph{Sequence Length Optimization}: We demonstrate that the FT measurement sequence length can be reduced to exactly $2r+1$—only one additional measurement beyond the original non-FT sequence. The consistent minimality across all computationally tractable cases, combined with the structural constraints derived in Theorem~\ref{th:FTMSC}, provides strong evidence for the conjectured optimality of this length.

Second, \emph{Theoretical Characterization}: (i) We establishes a complete characterization of syndrome matrices meeting distance-3 fault-tolerance requirements for general stabilizer codes (Theorem~\ref{th:FTMC}). (ii) We proves a necessary and sufficient condition for the existence of minimal-length ($2r+1$) FT sequences in quantum Hamming codes, ensuring full distinguishability between input errors and internal faults (Theorem~\ref{th:FTMSC}).

Last, \emph{Constructive Protocol}: We presents an explicit construction method using \textit{cyclic matrix operations} to generate these sequences through row combinations of initial stabilizers (Theorem~\ref{th:EoC}). The resulting sequences preserve a \textit{self-dual CSS-like symmetry}, enabling hardware-efficient circuit reuse via \textit{Pauli $X$-$Z$ exchange} between the first $r$ and subsequent $r$ stabilizer measurements, achieving a significant hardware overhead reduction.

For distance-3 FTQEC with quantum Hamming codes, our approach simultaneously reduces the time overhead via shorting the measurement sequence length and the hardware overhead through symmetry-enabled circuit multiplexing.
These results constitute a significant advancement in addressing a key open problem concerning minimal FT sequence design for quantum Hamming codes. Our symmetry-aware framework may further extend to other stabilizer codes, offering a systematic approach for overhead reduction in FT architectures.

\section{Preliminaries}

\subsection{Internal faults potentially confused with input errors}
We address a critical challenge in FT quantum computing: internal faults that either evade detection by conventional FT measurement circuits or require an impractically large number of measurement rounds. Such faults generate syndromes that are often indistinguishable from those caused by input errors, significantly complicating the error identification process and demanding an increased number of measurement rounds to accurately resolve syndrome patterns~\cite{N20}.
These internal faults mainly arise via two distinct pathways: first, faults occurring at the data-qubit terminals of two-qubit gates during data-ancilla interactions; second, propagating faults originating from measurement circuitry that induce single-error through backward propagation via two-qubit gate operations. Note that the cases inducing more errors through backward propagation can typically be captured by a well designed flag-based measurement circuit.

\begin{figure}[!t]
\centering
\includegraphics[width=2.5in]{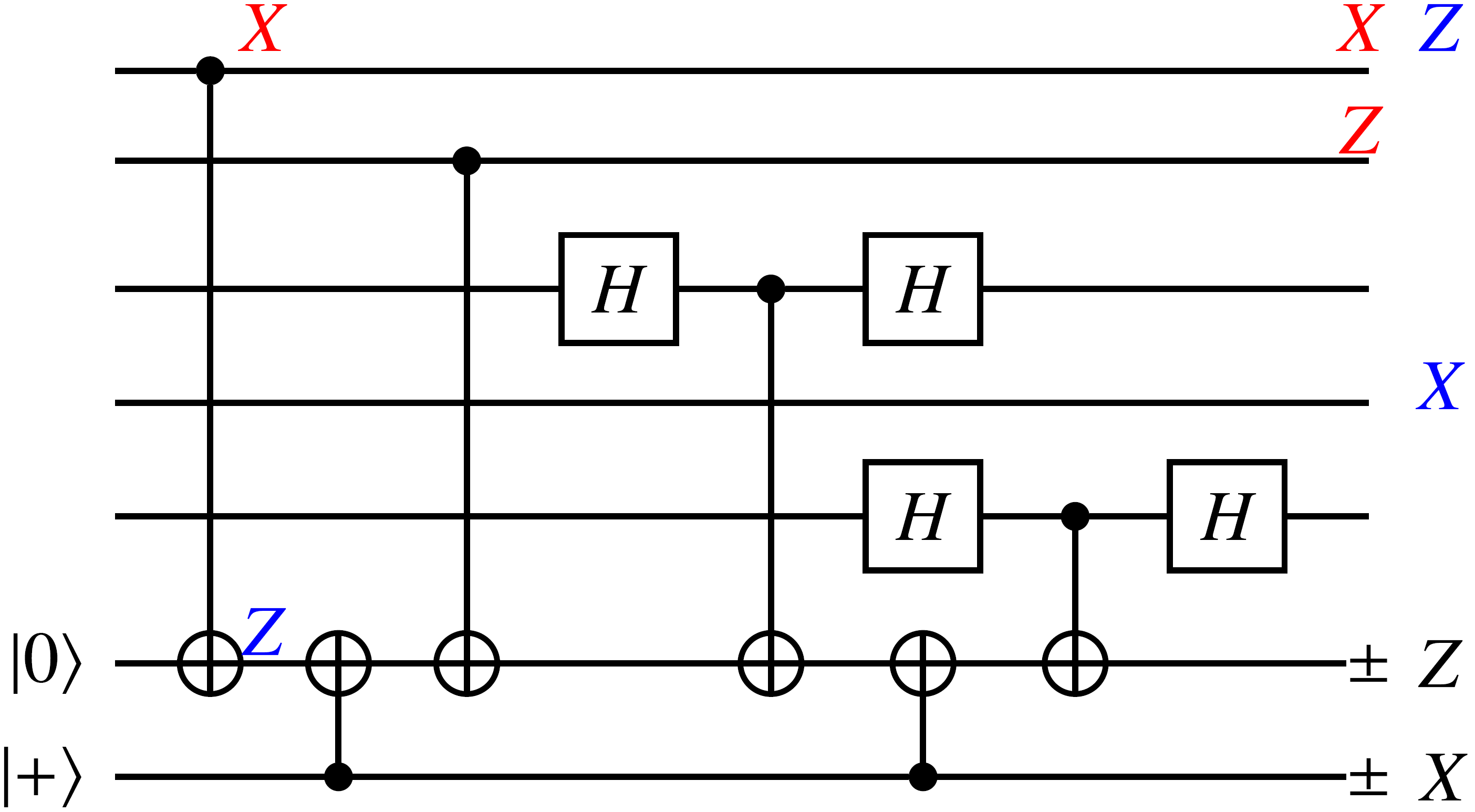}
\caption{Two representative error mechanisms arising from internal faults during the measurement of $ZZXIX$ using flag-based method for 5-qubit code. (Red): The errors caused by $X_1$ occurring on the control qubit during the operation of the first CNOT gate; (Blue): The errors caused by $Z$ occurring on the target (ancilla) qubit during the first CNOT gate operation. Here, black $X$ and $Z$ symbols denote measurement bases, while colored symbols indicate errors.}
\label{fig:internal faults}
\end{figure}

To illustrate the two error pathways, we analyze the five-qubit code $[\![5, 1, 3]\!]$~\cite{Laf96} with stabilizers $\langle XZZXI, ZZXIX, ZXIXZ, XIXZZ \rangle$. Suppose the first stabilizer $XZZXI$ is measured via a flag-based circuit, followed by $ZZXIX$ measurement (Fig.~\ref{fig:internal faults}).
If an $X$ error occurs on the controlled-NOT (CNOT) control qubit during the operation of the initial CNOT gate, and no subsequent faults happen, the syndrome $0010$ is observed. Notably, this syndrome is identical to that of an input $Z_{2}$ error under perfect extraction\footnote{Hereafter, $X_i$ ($Z_i$) denotes an $X$ (or $Z$) error on qubit $i$.}.
Critically, neither flag-based nor cat-state-based circuits can distinguish these cases. Without repeated extraction, the internal $X_1$ error remains uncorrected while a $Z_2$ error is spuriously introduced.

Furthermore, consider the case that a $Z$ error occurs on the target (ancilla) qubit during the first CNOT gate operation ( Fig.~\ref{fig:internal faults} blue). An equivalent error $Z_1$ (up to stabilizers), is generated.
The internal $Z_1$ error, in the absence of additional faults, produces the same syndrome ($0001$) as an input
$X_4$ error under ideal measurements. This syndrome collision results in the introduction of an additional $X_4$ error while the internal $Z_1$ error persists (up to stabilizers).

\subsection{Identifying internal faults with quantum Hamming codes using redundant stabilizers}
Hamming codes, introduced by Richard Hamming in 1950~\cite{Hamming50}, represent a foundational class of linear error-correcting codes that have profoundly influenced both classical and quantum information theory. Defined by the parameters $[2^{r}-1, 2^{r}-1-r, 3]$, these codes are uniquely characterized by their ability to detect and correct arbitrary single-bit errors. Furthermore, as perfect codes, Hamming codes achieve the theoretical Hamming bound~\cite{Lint98}, meaning their spheres of radius one around each codeword perfectly pack the entire space of possible received vectors without overlap. This optimality stems from their elegant algebraic construction, which leverages parity-check matrices to generate syndromes that uniquely identify error locations~\cite{Sloane77}.

In the realm of quantum error correction, Hamming codes have been extended to quantum Hamming codes, a family of CSS (Calderbank-Shor-Steane) stabilizer codes with parameters $[\![2^{r}-1, 2^{r}-1-2r, 3]\!]$~\cite{Shor961}. 
Quantum Hamming codes can correct arbitrary single-qubit errors and are optimal distance-3 codes~\cite{Gottesman97}, minimizing the physical qubit overhead required for FT quantum computation. 
One of the well-known quantum Hamming codes is Steane code $[\![7, 1, 3]\!]$ having stabilizers given by Eq.~(\ref{Eq:Stean Stabilizer}) typically.
\begin{equation}
\left(
\begin{array}{ccccccr}\label{Eq:Stean Stabilizer}
I & I & I & Z & Z & Z & Z\\
I & Z & Z & I & I & Z & Z\\
Z & I & Z & I & Z & I & Z\\
I & I & I & X & X & X & X\\
I & X & X & I & I & X & X\\
X & I & X & I & X & I & X
\end{array}\right)
\end{equation}

For an $[\![n,n-r_{Z}-r_{X},3]\!]$ CSS code with $r_{Z}$ independent $Z$-stabilizer generators $g_1, \cdots, g_{r_{Z}}$, FT $X$-error correction can be achieved by measuring all $r_{Z}$ generators sequentially, followed by measuring only $g_1, \cdots, g_{r_Z-1}$. This protocol ensures that the syndrome caused by an internal fault is distinguishable from the syndrome of an input error.
Similarly, FT $Z$-error correction requires measuring
$2r_{X}-1$ $X$-stabilizer generators in an analogous manner. Consequently, distance-3 FTQEC can be implemented for any $[\![n,n-r_{Z}-r_{X},3]\!]$ CSS code by measuring a total of $2(r_{Z}+r_{X})-2$ generators. Specifically, for a quantum Hamming code $[\![2^{r}-1, 2^{r}-1-2r, 3]\!]$, this protocol necessitates $4r-2$ syndrome bit measurements to achieve distance-3 FTQEC.

If mixed $X$ and $Z$ error correction is permitted, the number of required syndrome bit measurements can be significantly reduced.
Delfosse et. al. demonstrated that for distance-3 FTQEC with the $[\![7,1,3]\!]$ and $[\![15,7,3]\!]$ quantum Hamming codes, only $7$ and $9$ (i.e., $2r+1$) stabilizer measurements are sufficient, respectively. However, for general quantum Hamming codes, it remains unclear whether $2r+1$ stabilizer measurements suffice for distance-3 FTQEC. Moreover, determining the minimal-length sequence of Pauli measurements required for distance-3 FTQEC with quantum Hamming codes remains an important open problem.

\section{Distance-3 FTQEC with quantum Hamming codes via one additional
measurement beyond the original non-FT sequence}
While no rigorous theoretical proof currently exists, extensive numerical evidence suggests that the minimal-length Pauli measurement sequence required for distance-3 FTQEC with quantum Hamming codes is $2r+1$. In this section, we prove that for any quantum Hamming code $[\![2^{r}-1, 2^{r}-1-2r, 3]\!]$, where $r=3k+1$ with $k$ being a positive integer, $2r+1$ stabilizer measurements suffice to achieve distance-3 FTQEC vs. the Shor method which requires up to $((d+1)/2)^{2}$ repetitions of the syndrome measurements (i.e. with a measurement of length up to $4\times 2r$). Moreover, we demonstrate that the syndrome caused by any internal fault can be perfectly distinguished from that of any input error.

\subsection{A necessary and sufficient condition}
For the $[\![7,1,3]\!]$ Steane code, the complete set of 21 weight-one error syndromes is fully characterized by the following syndrome matrix derived from Eq.~(\ref{Eq:Stean Stabilizer}). The matrix consists of seven consecutive
$6\times3$ blocks, where each block column-wise encodes the syndromes for $X$, $Z$ and $Y$ errors on a respective physical qubit. Specifically, the leftmost block’s three columns correspond to syndromes for $X$, $Z$ and $Y$ errors on the first qubit, with subsequent blocks following this pattern for qubits $2$ through $7$.
\begin{equation}\label{eq:Stean syndrome}
\mathbf{A} = \begin{pmatrix}
		000 & 000 & 000 & 101 & 101 & 101 & 101 \\
		000 & 101 & 101 & 000 & 000 & 101 & 101 \\
		101 & 000 & 101 & 000 & 101 & 000 & 101 \\
		000 & 000 & 000 & 011 & 011 & 011 & 011 \\
		000 & 011 & 011 & 000 & 000 & 011 & 011 \\
		011 & 000 & 011 & 000 & 011 & 000 & 011
	\end{pmatrix}
\end{equation}

The FT design of measurement sequences for stabilizer codes is realized through carefully constructed redundant combinations of stabilizer generators. This procedure is mathematically equivalent to performing \emph{restricted elementary row operations} on the syndrome matrix, which include (i) row permutations and (ii) row additions implemented via bitwise modulo-2 arithmetic, thereby generating an optimized parity-check configuration.
Building on prior results~\cite{N20}, we formalize these observations in the following theorem:
\begin{theorem}(FT Measurement Criteria)\label{th:FTMC}
The distinguishing capability between an input error and an internal fault in a distance-3 FT stabilizer code is mathematically equivalent to the syndrome matrix satisfying the following two conditions:
\begin{enumerate}
\item{All columns must be distinct.}
\item{No column's syndrome subsequence (beginning from its first non-zero element) may coincide with any same-position subsequence in other matrix blocks.}
\end{enumerate}
\end{theorem}
The first condition trivially guarantees that the stabilizer code maintains distance-3 protection, while the second condition ensures that no two-qubit pair constitutes a \emph{bad} set~\cite{N20}.


Analogous to Eq.~(\ref{eq:Stean syndrome}), the weight-one error syndrome matrix for $[\![2^{r}-1, 2^{r}-1-2r, 3]\!]$ Hamming codes admits a block-matrix representation of dimension $2r\times (3\times(2^{r}-1))$. Through the equivalence between row operations and left matrix multiplication, we define:
\begin{itemize}
  \item $A\in \mathbb{F}_{2}^{2r\times 3(2^{r}-1)}$: The original syndrome matrix of the quantum Hamming code, where each block of three columns corresponds to the $X$, $Z$ and $Y$ error syndromes for a physical qubit.
  \item $B\in \mathbb{F}_{2}^{(2r+1)\times 3(2^{r}-1)}$: The transformed syndrome matrix generated by applying restricted elementary row operations to $A$. This matrix defines the FT measurement sequence.
  \item $C\in \mathbb{F}_{2}^{(2r+1)\times 2r}$: The full-rank transformation matrix satisfying $B=CA$, where $C$ is constructed via row permutations and $\mathbb{F}_{2}$-linear combinations (bitwise XOR operations).
\end{itemize}

This formulation reduces the FT measurement sequence optimization to constructing a matrix $C$ such that $B$
satisfies the conditions of Theorem~\ref{th:FTMC}.
Under the minimal measurement sequence length constraint of $2r+1$ for distance-3 FTQEC with quantum Hamming codes, we establish following theorem.

\begin{theorem}(\emph{FT Measurement Sequence Construction}).\label{th:FTMSC}
Let $[\![2^{r}-1, 2^{r}-1-2r, 3]\!]$ be a quantum Hamming code with syndrome matrix $A\in \mathbb{F}_{2}^{2r\times 3(2^{r}-1)}$.
There exists a distance-3 FT measurement sequence of length $2r+1$ with guaranteed distinguishability between input errors and internal faults, if and only if there exists a transformation matrix $C\in \mathbb{F}_{2}^{(2r+1)\times 2r}$ yielding $B=CA$ that satisfies:
\begin{enumerate}
\item{\emph{Column distinction:} All columns of $B$ are distinct, i.e. $rank(B)=2r$.}
\item{\emph{Forward Consistency:} For any column $\mathrm{\textbf{b}}_j \in B$ where $b_{ij}=0$ ($1\le i\le r-1$), the binary subsequence $(b_{kj})^{2r+1}_{k=i^{*}}$ beginning at the first nonzero entry $b_{i^{*},j}=1$ must be positionally unique from columns in other blocks of $B$.}
\item{\emph{Backward Consistency:} For any column $\mathrm{\textbf{b}}_j \in B$, if there exist another column $\mathrm{\textbf{b}}_k$ in other blocks of $B$ where $(b_{m,j})^{2r+1}_{m=r}=(b_{m,k})^{2r+1}_{m=r}$, then their preceding subsequences $(b_{nj})^{r-1}_{n=i^{*}}$ and $(b_{nk})^{r-1}_{n=i^{*}}$ must be different, where $b_{i^{*},j}=1$ is the first nonzero entry of $\mathrm{\textbf{b}}_j$.}
 \end{enumerate}
\end{theorem} 
\begin{proof}
  We demonstrate that the resulting FT measurement sequence yielded from $B=CA$ satisfies the two requirements of Theorem~\ref{th:FTMC} through the following two-part analysis:

 \indent{\emph{Part 1: Error correction capacity (All columns are distinct)}}

  It is trivial to require all columns of $B$ are distinct. Specifically,
  since $C$ is constructed via admissible row operations (permutations and $\mathbb{F}_{2}$-linear combinations) and $rank(A) = 2r$ by construction of the quantum Hamming code, the product $B = CA$ with $rank(B)=2r$ ensures the row equivalence between $A$ and $B$, guaranteeing unique identification of any weight-1 input Pauli error ($X$, $Y$, or $Z$) on physical qubits.

 \indent{\emph{Part 2: Fault-Temporal Distinguishability (Consistency Conditions)}}

 We analyze the forward and backward consistency conditions by partitioning the columns of $B$ into two cases based on their leading $r-1$ entries:

 \emph{Case 1: } leading $r-1$ entries all zero.

 For any column $\mathrm{\textbf{b}}_j$ where $b_{ij}=0$ ($1\le i\le r-1$), let $b_{i^{*},j}=1$ be its first nonzero entry. The Forward Consistency condition requires the subsequence $(b_{i^{*},j}, \cdots, b_{2r+1,j})$ to be unique among all columns in other blocks of $B$. This condition directly implies that any internal fault occurring during the measurement sequence will alter this subsequence, producing a syndrome distinguishable from input errors occurring in other blocks.

  \emph{Case 2: } leading $r-1$ entries not all zero.

  For columns with $b_{i^{*},j}=1$ ($i^{*}\le r-1$) being the first nonzero entry, we further consider:

  \indent 1. If the trailing $r+2$ entries $(b_{r,j},\cdots, b_{2r+1, j})$ differ from all corresponding subsequences in other blocks, the uniqueness of the extended subsequence $(b_{i^{*},j},\cdots, b_{2r+1, j})$ trivially satisfies the second condition of Theorem~\ref{th:FTMC}.

  \indent 2. If another column $\textbf{b}_{k}$ shares identical trailing $r+2$ entries, the Backward Consistency condition mandates that their preceding subsequences $(b_{i^{*},j},\cdots, b_{r-1, j})$ and $(b_{i^{*},k},\cdots, b_{r-1, k})$ must differ. This ensures the second condition of Theorem~\ref{th:FTMC} is satisfied.

  The full column rank guarantees single-error correction, while the consistency conditions enforce a temporal separation between input errors and internal faults. Thereby, $B$ defines a FT measurement sequence of length $2r+1$ meeting the requirements of Theorem~\ref{th:FTMC}.
\end{proof}

\subsection{Construction of the Cyclic Transformation Matrices}
Building upon Theorem~\ref{th:FTMSC}, we present our main contribution: an explicit construction of the transformation matrix $C$ for the family of quantum Hamming codes with parameters $r\equiv 1~mod~3$. This constructive approach reveals an interesting cyclic symmetry in the FT syndrome extraction mechanisms of quantum Hamming codes.
\begin{theorem}(\emph{Constructive Existence of FT Cyclic Extension}).\label{th:EoC}
For any quantum Hamming code $[\![2^{r}-1, 2^{r}-1-2r, 3]\!]$ with $r=3k+1 (k\in \mathbb{Z}^{+})$ and syndrome matrix $A\in \mathbb{F}_{2}^{2r\times 3(2^{r}-1)}$, there exists an explicit cyclic transformation matrix $C\in \mathbb{F}_{2}^{(2r+1)\times 2r}$ generated by the polynomial $g(x)=1+x^{r+1}+x^{2r-1}$ under modulo $x^{2r}-1$ arithmetic, such that the product $B=CA$ corresponds to a distance-3 FT measurement sequence of length $2r + 1$ with guaranteed
distinguishability between input errors and internal faults.
\end{theorem}
The complete proof appears in Appendix A, where we verify the resulting $B$ satisfies all conditions of Theorem~\ref{th:FTMSC} for FT measurement sequences. To highlight the practical implications of Theorem~\ref{th:EoC}, we present a concrete example below.

\
\begin{example} \label{ex:FT-sequence}
For the quantum Hamming code $[\![15, 7, 3]\!]$ with parameter $r=4$,
Theorem~\ref{th:EoC} guarantees the existence of a cyclic transformation matrix $C$ generated by the polynomial $g(x)=1+x^{5}+x^{7}$. The explicit form of $C\in \mathbb{F}^{9\times 8}_{2}$ is:
\begin{equation}
\mathbf{C} = \left(
\begin{array}{cccccccc}
	1 & 0 & 0 & 0 & 0 & 1 & 0 & 1 \\
    1 & 1 & 0 & 0 & 0 & 0 & 1 & 0 \\
    0 & 1 & 1 & 0 & 0 & 0 & 0 & 1 \\
    1 & 0 & 1 & 1 & 0 & 0 & 0 & 0 \\
    0 & 1 & 0 & 1 & 1 & 0 & 0 & 0 \\
    0 & 0 & 1 & 0 & 1 & 1 & 0 & 0 \\
    0 & 0 & 0 & 1 & 0 & 1 & 1 & 0 \\
    0 & 0 & 0 & 0 & 1 & 0 & 1 & 1 \\
    1 & 0 & 0 & 0 & 0 & 1 & 0 & 1
\end{array}
\right)
\end{equation}

The standard stabilizer generators of the $[\![15, 7, 3]\!]$ code are represented as:
\begin{equation}
\setlength{\arraycolsep}{2.5pt}
\left(
\begin{array}{ccccccccccccccr}\label{Eq:15 Stabilizer}
I & I & I & I & I & I & I & Z & Z & Z & Z & Z & Z & Z & Z\\
I & I & I & Z & Z & Z & Z & I & I & I & I & Z & Z & Z & Z\\
I & Z & Z & I & I & Z & Z & I & I & Z & Z & I & I & Z & Z\\
Z & I & Z & I & Z & I & Z & I & Z & I & Z & I & Z & I & Z\\
I & I & I & I & I & I & I & X & X & X & X & X & X & X & X\\
I & I & I & X & X & X & X & I & I & I & I & X & X & X & X\\
I & X & X & I & I & X & X & I & I & X & X & I & I & X & X\\
X & I & X & I & X & I & X & I & X & I & X & I & X & I & X
\end{array}\right),
\end{equation}

The corresponding syndrome matrix $A\in \mathbb{F}^{8\times 45}_{2}$ (with each $3$-bit block representing
$X/Z/Y$ error syndromes) is:
\begin{equation}
\setlength{\arraycolsep}{-0.5pt}
\left(
\begin{array}{ccccccccccccccc}
   000 & 000 & 000 & 000 & 000 & 000 & 000 & 101 & 101 & 101 & 101 & 101 & 101 & 101 & 101 \\
   000 & 000 & 000 & 101 & 101 & 101 & 101 & 000 & 000 & 000 & 000 & 101 & 101 & 101 & 101 \\
   000 & 101 & 101 & 000 & 000 & 101 & 101 & 000 & 000 & 101 & 101 & 000 & 000 & 101 & 101 \\
   101 & 000 & 101 & 000 & 101 & 000 & 101 & 000 & 101 & 000 & 101 & 000 & 101 & 000 & 101 \\
   000 & 000 & 000 & 000 & 000 & 000 & 000 & 011 & 011 & 011 & 011 & 011 & 011 & 011 & 011 \\
   000 & 000 & 000 & 011 & 011 & 011 & 011 & 000 & 000 & 000 & 000 & 011 & 011 & 011 & 011 \\
   000 & 011 & 011 & 000 & 000 & 011 & 011 & 000 & 000 & 011 & 011 & 000 & 000 & 011 & 011 \\
   011 & 000 & 011 & 000 & 011 & 000 & 011 & 000 & 011 & 000 & 011 & 000 & 011 & 000 & 011
 \end{array}
  \right).
 \end{equation}
 Thus, the transformed syndrome matrix $B$ can be obtained by the production of $B=CA \in \mathbb{F}^{9\times 45}_{2}$ as follows,
 \begin{equation}\label{Eq:15B}
\setlength{\arraycolsep}{-0.5pt}
  \left(
\begin{array}{ccccccccccccccc}
   011 & 000 & 011 & 011 & 000 & 011 & 000 & 101 & 110 & 101 & 110 & 110 & 101 & 110 & 101 \\
   000 & 011 & 011 & 101 & 101 & 110 & 110 & 101 & 101 & 110 & 110 & 000 & 000 & 011 & 011 \\
   011 & 101 & 110 & 101 & 110 & 000 & 011 & 000 & 011 & 101 & 110 & 101 & 110 & 000 & 011 \\
   101 & 101 & 000 & 000 & 101 & 101 & 000 & 101 & 000 & 000 & 101 & 101 & 000 & 000 & 101 \\
   101 & 000 & 101 & 101 & 000 & 101 & 000 & 011 & 110 & 011 & 110 & 110 & 011 & 110 & 011 \\
   000 & 101 & 101 & 011 & 011 & 110 & 110 & 011 & 011 & 110 & 110 & 000 & 000 & 101 & 101 \\
   101 & 011 & 110 & 011 & 110 & 000 & 101 & 000 & 101 & 011 & 110 & 011 & 110 & 000 & 101 \\
   011 & 011 & 000 & 000 & 011 & 011 & 000 & 011 & 000 & 000 & 011 & 011 & 000 & 000 & 011 \\
   011 & 000 & 011 & 011 & 000 & 011 & 000 & 101 & 110 & 101 & 110 & 110 & 101 & 110 & 101
 \end{array}
  \right).
 \end{equation}
 Correspondingly, we have the following resulted FT stabilizer measurement sequence,
 \begin{equation}\label{Eq:15FTMC}
\setlength{\arraycolsep}{2.5pt}
\left(
\begin{array}{ccccccccccccccc}
   X & I & X & X & I & X & I & Z & Y & Z & Y & Y & Z & Y & Z \\
   I & X & X & Z & Z & Y & Y & Z & Z & Y & Y & I & I & X & X \\
   X & Z & Y & Z & Y & I & X & I & X & Z & Y & Z & Y & I & X \\
   Z & Z & I & I & Z & Z & I & Z & I & I & Z & Z & I & I & Z \\
   Z & I & Z & Z & I & Z & I & X & Y & X & Y & Y & X & Y & X \\
   I & Z & Z & X & X & Y & Y & X & X & Y & Y & I & I & Z & Z \\
   Z & X & Y & X & Y & I & Z & I & Z & X & Y & X & Y & I & Z \\
   X & X & I & I & X & X & I & X & I & I & X & X & I & I & X \\
   X & I & X & X & I & X & I & Z & Y & Z & Y & Y & Z & Y & Z
\end{array}
\right).
\end{equation}

\end{example}
One can verify that the transformed syndrome matrix $B$ (Eq.~\ref{Eq:15B}) satisfies all conditions of Theorem~\ref{th:FTMC}, demonstrating that the derived measurement sequence of length 9 (Eq.~\ref{Eq:15FTMC}) is distance-3 FT with guaranteed distinguishability between input errors and internal faults. In contrast, Shor's method requires up to 4 rounds of measurements of the original stabilizer sequence, potentially resulting in a total measurement sequence length of 32. A short sequence of the same length was also found in Ref.~\cite{N20}, while our method offers a systematical way to construct the optimized measurement sequence and enjoys a cyclic symmetry.
As is straightforward to observe that, the proposed FT measurement sequence (Eq.~\ref{Eq:15FTMC}) exhibits a distinct symmetric property-the subsequent $4$ rows can be obtained by interchanging $Z$ and $X$ in the first $4$ rows, with the final row being a repetition of the first row.
\section{Discussion}
The syndrome matrix $B$ derived from our cyclic transformation matrix $C$ satisfies an even more stringent property: no column's syndrome subsequence (starting from its first non-zero element) may coincide with any same-position subsequence within \emph{the entire matrix B, including subsequences within its own block}. While the previously stated condition (distinctness across different blocks) already guarantees distance-3 fault tolerance, our construction provides a stronger guarantee for error correction confidence.

For instance, in Ref.~\cite{N20}, Delfosse and Reichardt present a length-6 FT measurement sequence for the 5-qubit code (Eq.~(\ref{Eq:5FTMC})) and its syndrome matrix  (Eq.~(\ref{Eq:5FTMCM})), respectively.
 \begin{equation}\label{Eq:5FTMC}
\setlength{\arraycolsep}{2.5pt}
\left(
\begin{array}{ccccc}
  X & Z & Z & X & I \\
  Z & Y & Y & Z & I \\
  I & X & Z & Z & X \\
  X & I & X & Z & Z \\
  Z & Y & Y & Z & I \\
  X & Z & Z & X & I
\end{array}
\right)
\end{equation}

 \begin{equation}\label{Eq:5FTMCM}
\setlength{\arraycolsep}{1.1pt}
\left(
\begin{array}{ccccccccccccccc}
  \underline{X_{1}} & \underline{Z_{1}} & \underline{Y_{1}} & ~\underline{X_{2}} & \underline{Z_{2}} &  \underline{Y_{2}} & ~\underline{X_{3}} & \underline{Z_{3}} & \underline{Y_{3}} & ~\underline{X_{4}} &\underline{Z_{4}} & \underline{Y_{4}} & ~\underline{X_{5}} & \underline{Z_{5}} & \underline{Y_{5}} \\
  0 & 1 & 1 & ~ 1& 0& 1&~ 1& 0 & 1& ~0& 1 & 1& ~0 & 0 & 0 \\
  1 & 0 & 1 & ~ 1& 1& 0&~ 1& 1 & 0& ~1& 0 & 1& ~0 & 0 & 0 \\
  0 & 0 & 0 & ~ 0& 1& 1&~ 1& 0 & 1& ~1& 0 & 1& ~0 & 1 & 1 \\
  0 & 1 & 1 & ~ 0& 0& 0&~ 0& 1 & 1& ~1& 0 & 1& ~1 & 0 & 1 \\
  1 & 0 & 1 & ~ 1& 1& 0&~ 1& 1 & 0& ~1& 0 & 1& ~0 & 0 & 0 \\
  0 & 1 & 1 & ~ 1& 0& 1&~ 1& 0 & 1& ~0& 1 & 1& ~0 & 0 & 0
\end{array}
\right)
\end{equation}
One can verify that when a $Y$ error occurs on the fifth qubit between the third and fourth stabilizer measurements, the syndrome pattern ``000100" coincides exactly with that of an $X$ input error on the same qubit. Consequently, the decoder would misinterpret this internal $Y$ error as an $X$ input error on the fifth qubit. While this does not introduce additional errors, the original $Y$ error remains uncorrected in that cycle. This ambiguity arises precisely because the construction only enforces subsequence distinctness across different blocks but allows overlaps within the same block.

Moreover, as illustrated in example 4, our construction generates FT measurement sequences for quantum Hamming codes that inherit remarkable symmetry properties from the code structure.
The characteristic $X-Z$ symmetry in the stabilizer generators of conventional quantum Hamming codes, combined with the circulant structure of the C-matrix, guarantees that the first $r$ rows and subsequent $r$ rows of the derived FT stabilizer measurement sequences exhibit identical configurations up to Pauli operator permutations.

Specifically, the following $r$ rows are obtained from the initial $r$ rows through a permutation of Pauli $X$ and $Z$ operators (note that $Y = iXZ$ is preserved under this transformation, ignoring global phases), with the final row completing the cycle by reproducing the first row. This symmetric structure is explicitly demonstrated in Eq.~(\ref{Eq:15FTMC}), where the sequential three rows are generated from the first three rows by systematic $X-Z$ exchange.

The self-dual CSS-like symmetry exhibited in our FT measurement sequences enables hardware reuse in quantum stabilizer circuits. Specifically, the measurement circuits for the subsequent $r$ stabilizers can be obtained by simply adding Hadamard ($H$) gates at both ends of the circuits implementing the first $r$ stabilizers, eliminating the need for additional quantum circuitry.

This symmetry suggests an efficient hardware multiplexing strategy: by designing a switching system to collectively control these $H$ gates, we can disable them during measurements of the first $r$ stabilizers and activate them for the remaining $r$ stabilizers. Such circuit reuse significantly reduces the hardware overhead of FT measurement schemes.

\section{Conclusion}
To address the critical challenge of high overhead in FTQEC, we have developed a systematic framework for constructing efficient measurement sequences for quantum Hamming codes $[\![2^r-1, 2^r-1-2r, 3]\!]$ where $r=3k+1$ ($k \in \mathbb{Z}^+$). Our work specifically tackles the important problem of internal faults that mimic input errors and become amplified through non-FT correction cycles.

We prove that distance-3 FTQEC can be achieved with just $2r+1$ measurements - only one additional measurement beyond the original non-FT sequence. Exhaustive verification confirms this represents the minimal length for these cases.
We have provided a complete characterization of syndrome matrices meeting distance-3 FT requirements (Theorem 1) and established a necessary and sufficient condition for the existence of length-$(2r+1)$ FT sequences with guaranteed distinguishability between input errors and internal faults (Theorem 2). We also have developed an explicit method to generate these sequences while preserving the self-dual CSS-like symmetry (Theorem 3). This symmetry enables hardware-efficient circuit reuse via Pauli $X-Z$ exchange.

Exhaustive search confirms that for quantum Hamming codes mentioned in this paper, the minimal FT sequence length is indeed $2r+1$, demonstrating that our construction achieves optimality for these cases. While we conjecture that $2r+1$ represents the minimal sequence length for general distance-3 quantum Hamming codes, a formal proof remains an important open challenge - a significant step toward which has been established in this work.

Furthermore, our current cyclic construction is explicitly developed for codes with $r = 3k+1 (k \in \mathbb{Z}^+)$. The extension to cases where $r = 3k$ or $r = 3k+2$ requires additional investigation and represents an important direction for future research. We leave this generalization to subsequent work, along with the development of a unified framework for all quantum Hamming code families.

This work provides both theoretical insights and practical tools for optimizing FT quantum computation, while establishing a foundation for further investigations into resource-efficient quantum error correction.


\section{Appendix: Proof of Theorem 3}
For convenience, we partition matrix $C$ into four submatrices. The first $r$ columns form the left part, while the remaining $r$ columns constitute the right part; the first $r-1$ rows make up the upper part, and the subsequent $r+2$ rows form the lower part. The resulting four submatrices are denoted as $C_{11}, C_{12}, C_{21},$ and $C_{22}$, which are of size $(r-1)\times r, (r-1)\times r, (r+2)\times r$, and  $(r+2)\times r$, respectively, forming a matrix
\begin{eqnarray}
\mathbf{C}&=&
\setlength{\arraycolsep}{2.2pt}
\left(
\begin{array}{cccccc|cccccc}
	1 & 0 & 0 & \cdots & 0 & 0 & 0 & 1 & 0 & \cdots & 0 & 1 \\
	1 & 1 & 0 & \cdots & 0 & 0 & 0 & 0 & 1 & 0 & \cdots & 0 \\
	\vdots & \ddots & \ddots & \ddots & \ddots & \ddots & \ddots & \ddots & \ddots & \ddots & \ddots & \vdots \\
	0 & \cdots & 0 & 1 & 1 & 0 & 0 & 0 & 0 & \cdots & 0 & 1 \\ \hline
	1 & 0 & \cdots & 0 & 1 & 1 & 0 & 0 & 0 & \cdots & 0  & 0 \\
    0 & 1 & 0 & \cdots & 0 & 1 & 1 & 0 & 0 & \cdots & 0  & 0 \\
	\vdots & \ddots & \ddots & \ddots & \ddots & \ddots & \ddots & \ddots & \ddots & \ddots & \ddots & \vdots \\
    0 & 0 & \cdots & 0 & 0 & 1 & 0 & 0 & \cdots & 1 & 1 & 0\\
	0 & 0 & \cdots & 0 & 0 & 0 & 1 & 0 & \cdots & 0 & 1 & 1\\
    1 & 0 & 0 & \cdots & 0 & 0 & 0 & 1 & 0 & \cdots & 0 & 1
\end{array}
\right)
\nonumber \\
&=&\left(
\begin{array}{c|c}
	\mathbf{C_{11}} & \mathbf{C_{12}} \\ \hline
	\mathbf{C_{21}} & \mathbf{C_{22}}
\end{array}
\right).\label{splitted}
\end{eqnarray}

On the other hand, the columns of $A$ can be divided into the following three types
\begin{equation}\label{A-3-types}
\begin{pmatrix}
	\boldsymbol{v}\\
	0
\end{pmatrix},
\begin{pmatrix}
	0\\
	\boldsymbol{v}
\end{pmatrix},
\begin{pmatrix}
	\boldsymbol{v} \\
	\boldsymbol{v}
\end{pmatrix},
\end{equation}
corresponding to the syndromes of $X$, $Z$ and $Y$ errors, respectively, where $\boldsymbol{v} \in V_r =\{(a_1, a_2, \cdots, a_r)^{T}| a_i \in {0, 1}, \boldsymbol{v}\neq 0\}$. According to the perfect CSS structure of quantum Hamming codes, $\boldsymbol{v}$ precisely covers the entire space of non-zero binary sequences of length $r$. %

All columns of the resulted $B=CA$ then can be divided into the following 3 types correspondingly,
\begin{equation}\label{B-3-types}
\begin{pmatrix}
	\mathbf{C_{11}} \cdot \boldsymbol{v}\\
	\mathbf{C_{21}} \cdot \boldsymbol{v}
\end{pmatrix},
\begin{pmatrix}
	\mathbf{C_{12}} \cdot \boldsymbol{v}\\
	\mathbf{C_{22}} \cdot \boldsymbol{v}
\end{pmatrix},
\begin{pmatrix}
	(\mathbf{C_{11}} + \mathbf{C_{12}}) \cdot \boldsymbol{v} \\
	(\mathbf{C_{21}} + \mathbf{C_{22}}) \cdot \boldsymbol{v}
\end{pmatrix}.
\end{equation}
In the following, we will show $B$ satisfies all the three requirements proposed in Theorem 2.

\subsection{The proof of all columns of $B$ are distinct}
  We begin this proof by proving $C$ is full columns rank, i.e $rank(C)=2r$, where $C$ is generated by the polynomial $g(x)=1+x^{r+1}+x^{2r-1}$ under modulo $f(x)=x^{2r}-1$ with $r=3k+1$.  

\begin{proof}
The roots of \( f(x) \) are the \( 2r \)-th roots of unity:
\[
\omega_m = e^{2\pi i m / 2r}, \quad m = 0, 1, \dots, 2r - 1.
\]
To prove coprimality, we show \( f(\omega_m) \neq 0 \) for all \( m \).
For \( x = \omega_m \), \( x^{2r} = 1 \) implies \( x^{2r-1} = x^{-1} \). Thus:
\[
g(x) = 1 + x^{r+1} + x^{-1}.
\]
Multiplying by \( x \) yields:
\begin{equation}
x + x^{r+2} + 1 = 0. \label{eq:simplified}
\end{equation}
Since \( x^{2r} = 1 \), \( x^r = \pm 1 \).

\textbf{Case 1}: \( x^r = 1 \)
Substituting into \eqref{eq:simplified}:
\[
x + x^2 + 1 = 0 \implies x^2 + x + 1 = 0.
\]
The roots are \( \zeta = e^{2\pi i/3} \) and \( \zeta^2 \). For \( r = 3k + 1 \):
\[
\zeta^{6k + 2} = \zeta^{4\pi i (3k + 1)/3} = \zeta^{4\pi i /3} \neq 1.
\]
No valid common roots exist.

\textbf{Case 2}: \( x^r = -1 \)
Substituting into \eqref{eq:simplified}:
\[
x - x^2 + 1 = 0 \implies x^2 - x - 1 = 0.
\]
The roots \( x = (1 \pm \sqrt{5})/2 \) lie off the unit circle (\( |x| \neq 1 \)), hence invalid.

Thus, $g(x)$ and $f(x)$ are coprime, and $C$ is full column rank, i.e. $rank(C)=2r$. Considering all columns of $A$ are distinct, we then can derive that all columns of $B=CA$ are distinct and $rank(B)=2r$.
\end{proof}

\subsection{The proof of forward consistency}
\begin{proof}
The first $r-1$ rows of the matrix $B$ can be categorized into the following classes: $C_{11} \cdot v$, $C_{12} \cdot v$ and $(C_{11}+C_{12}) \cdot v$.

We first examine the case where $C_{11} \cdot v =\textbf{0}$. This equality is satisfied if and only if $v=v_{i}=(00\cdots 01)^{T}$. 
Consequently, the subsequence comprising the last $r+2$ entries of this column (with the first $r-1$ elements being zero) is given by $C_{21}\cdot v_{i}=(110\cdots 0100)^{T}$.
We now prove that no other column in $B$ admits a subsequence (formed by its last $r+2$ entries) identical to $C_{21}\cdot v_{i}$. This follows from three observations:
\begin{enumerate}
\item{\emph{Uniqueness under $C_{21}$:} Since $C_{21}$ has full column rank, no other vector $v'\neq v_{i}$ yields $C_{21}\cdot v' = C_{21}\cdot v_{i}$.}
\item{\emph{Leading zeros in $C_{22}\cdot v$:} All subsequences generated by $C_{22}\cdot v$ begin with a $0$, whereas $C_{21}\cdot v_{i}$ starts with $1$, rendering them distinct.}
\item{\emph{Sturcture of $C_{21}+C_{22}$:} The $(r+2)\times r$ matrix
\begin{equation}
\mathbf{C_{21}+C_{22}} = \left(
\begin{array}{ccccccc}
	1 & 0 & \cdots & 0  & 0 & 1 & 1 \\
	1 & 1 & 0 & \cdots & 0 & 0 & 1 \\
	1 & 1 & 1 & 0 & \cdots  & 0 & 0 \\
	\vdots & \ddots & \ddots & \ddots & \ddots & \ddots & \vdots \\
	0 & \cdots & 0 & 0 & 1 & 1 & 1 \\
    1 & 0 & \cdots & 0 & 0 & 1 & 1 \\
	1 & 1 & 0 & \cdots & 0 & 0 & 1
\end{array}
\right)
\end{equation}

exhibits identical first two and last two rows. However, the corresponding rows of $C_{21}\cdot v_{i}=(110\cdots 0100)^{T}$ are opposite in parity. Thus, no $(C_{21}+C_{22}) \cdot v$ can match $C_{21}\cdot v_{i}$.}
\end{enumerate}
\emph{Conclusion}: for the case $C_{11} \cdot v =\textbf{0}$, the forward consistency condition holds.

We then consider the case where $C_{12} \cdot v =\textbf{0}$, which holds if and only if $v=v_{j}=(10\cdots00)^{T}$. The resulting subvector formed by the last $r+2$ entries of this column is $C_{22}\cdot v_{j}=(010\cdots 010)^{T}$. Notably, $C_{22}\cdot v_{j}$ begins with $01$, we proceed to demonstrate that the $r+1$-length subsequence starting from the second bit $1$ is unique among all columns. This uniqueness follows from three key observations:
\begin{enumerate}
\item{\emph{Uniqueness under $C_{22}$:} Due to the full column rank of $C_{22}$, there exists no $v'\neq v_{j}$ such that $C_{22}\cdot v' = C_{22}\cdot v_{j}$.}
\item{\emph{Distinctness from $C_{21}$ outputs:} All columns in $C_{21}$ have $0$ as their penultimate bit, while $C_{22}\cdot v_{j}$ has $1$ in this position. Consequently, no  $v_{i}$ satisfies $C_{21}\cdot v_{i} = C_{22}\cdot v_{j}$.}
\item{\emph{Non-match with $(C_{21}+C_{22})$ outputs:} 
    The antipodal symmetry between the first two and last two elements of $C_{22}\cdot v_{j}$ precludes any
    $(C_{21}+C_{22}) \cdot v$ from matching $C_{22}\cdot v_{j}$.}
\end{enumerate}
\emph{Conclusion}: for the case $C_{12} \cdot v =\textbf{0}$, the forward consistency condition holds.

Finally, we examine the case where $(C_{11}+C_{12}) \cdot v = \textbf{0}$. The matrix $(C_{11}+C_{12})$ has dimensions $(r-1)\times r$ and takes the following form:
\begin{equation}
\mathbf{C_{11}+C_{12}} =
\left(
\begin{array}{ccccccc}
1 & 1 & 0 & \cdots & 0 & 0 & 1 \\
1 & 1 & 1 & 0 & \cdots & 0 & 0 \\
0 & 1 & 1 & 1 & \cdots & 0 & 0 \\
\vdots & \ddots & \ddots & \ddots & \ddots & \ddots & \vdots \\
0 & \cdots & 1 & 1 & 1 & 0 & 0 \\
0 & 0 & \cdots & 1 & 1 & 1 & 0 \\
0 & 0 & 0 & \cdots & 1 & 1 & 1
\end{array}
\right)
\end{equation}
Notably, Rank$(C_{11}+C_{12}) = r-1$, implying a unique non-zero solution exists modulo 2. Through structural analysis, we find that $(C_{11}+C_{12}) \cdot v_{l} = \textbf{0}$ is satisfied by a vector $v_{l}$ of length $r$ consisting of $k$ repetitions of 101 followed by a final 1 (where $r=3k+1$ for positive integer $k$), i.e. $v_{l} = (101\cdots 1011)$. This solution yields $(C_{21}+C_{22}) \cdot v_{l} = v_{l}'$, where $v_{l}'=(10\underbrace{00\cdots00}_{2+3(k-1)}10)$. We now prove that no other column in B contains a subvector (formed by its last $r+2$ elements) identical to $v_{l}'$:
\begin{enumerate}
\item{\emph{Distinctness from $C_{21}$ outputs:} The penultimate element of $v_{l}'$ is 1, while all columns of $C_{21}\cdot v$ have 0 in their penultimate position, making them distinct.}
\item{\emph{Distinctness from $C_{22}$ outputs:} The first element of $v_{l}'$ is 1, whereas all columns of $C_{22}\cdot v$ begin with 0, ensuring distinction.}
\item{\emph{Uniqueness under $(C_{21}+C_{22})$:} Since ($C_{21}+C_{22}$) has full column rank, no other vector $v_{l'} \neq v_{l}$ satisfies $(C_{21}+C_{22})\cdot v_{l'} = (C_{21}+C_{22})\cdot v_{l}$}
\end{enumerate}
\emph{Conclusion}: for the case  $(C_{11}+C_{12}) \cdot v = \textbf{0}$, the forward consistency condition holds.
\end{proof}

\subsection{The proof of backward consistency}
If a column in $B$ where the first $r-1$ elements are not all zero and the subvector formed by its last $r+2$ elements (generated by $C_{21}$, $C_{22}$, or $C_{21}+C_{22}$) matches the corresponding subvector of another column in $B$. Then exactly one of the following cases must hold.

\emph{Case 1: Same Matrix, Different Vectors.} This scenario involves collisions where the same matrix multiplies two distinct vectors to produce identical subvectors. Specifically, there exist $v_i \neq v_j$: $C_{21} \cdot v_i = C_{21} \cdot v_j, C_{22} \cdot v_i = C_{22} \cdot v_j, (C_{21} + C_{22}) \cdot v_i = (C_{21} + C_{22}) \cdot v_j$. However, since $C_{21}$, $C_{22}$ and $(C_{21} + C_{22})$ all have full column rank, these equalities require $v_i = v_j$, contradicting the distinctness assumption.

\emph{Case 2: Different Matrices, Same Vector.} Here, we examine whether a nonzero single vector $v_i$ can produce identical subvectors under different matrix multiplications. For $C_{21}\cdot v_i = C_{22}\cdot v_i$, we derive $(C_{21}-C_{22})\cdot v_i = 0$. However, $(C_{21}-C_{22})$, same as the matrix $(C_{21}+C_{22})$, is of full column rank, there is no nonzero vector ensure this equation holds. For the other two equalities $C_{21}\cdot v_i = (C_{21} + C_{22}) \cdot v_i$ and $C_{22}\cdot v_i = (C_{21} + C_{22}) \cdot v_i$, they simplify to $C_{22}\cdot v_i$ and $C_{21}\cdot v_i$, respectively, contradicting the full column rank character of $C_{21}$ and $C_{22}$.

\emph{Case 3: Different Matrices, Different Vector.} Suppose there exist non-zero, distinct vectors $v_i, v_j \in V_r$, such that $C_{21}\cdot v_i = C_{22}\cdot v_j$. Let $v_i = (a_1, a_2, \cdots, a_r)^T, v_j = (b_1, b_2, \cdots, b_r)^T$, This implies the derived relations $(C_{21} + C_{22}) \cdot v_i = C_{22} \cdot (v_i + v_j), (C_{21} + C_{22}) \cdot v_j = C_{21} \cdot (v_i + v_j)$. Therefor, it is sufficient to just focus on the case $C_{21}\cdot v_i = C_{22}\cdot v_j$.
Recall the structure of $C_{21}$ and $C_{22}$, we have $v_i = T\cdot v_j$, where the submatrix formed by rows $3$ through $r$ of matrix $T$ constitutes a circulant matrix of dimensions $(r-2)\times r$, generated by the polynomial $1 + x$, i.e.
\begin{equation}\label{abrelations}
\begin{pmatrix}
	a_1\\
    a_2\\
    a_3\\
    \vdots\\
    a_{r-2}\\
    a_{r-1}\\
    a_r
\end{pmatrix}
 = \begin{pmatrix}
	0 & 1 & 0 & \cdots & 0 & 0 & 1 \\
	1 & 0 & 0 & \cdots & 1 & 1 & 0 \\
	1 & 1 & 0 & \cdots & 0 & 0 & 0 \\
	0 & 1 & 1 & \cdots & 0 & 0 & 0 \\
	\vdots & \ddots & \ddots & \ddots & \ddots & \ddots & \vdots \\
	0 & 0 & 0 & \cdots & 1 & 1 & 0
\end{pmatrix} \cdot
\begin{pmatrix}
	b_1\\
    b_2\\
    b_3\\
    \vdots\\
    b_{r-2}\\
    b_{r-1}\\
    b_r
\end{pmatrix}
\end{equation}
Consequently, for the corresponding first $r-1$ rows, we have the following equations,
\begin{subequations}\label{C11-C12vb}
\begin{align}
  \mathbf{C_{11}} \cdot \boldsymbol{v_i} = \mathbf{C_{11}} \cdot \mathbf{T} \cdot \boldsymbol{v_j} &= \begin{pmatrix}
	b_2 + b_r \\
	b_2 + b_{r-2} \\
	b_2 + b_{r-2} + b_{r-1} \\
	b_1 + b_3 \\
	b_2 + b_4 \\
	\vdots \\
    b_{r-5} + b_{r-3}\\
	b_{r-4} + b_{r-2}
\end{pmatrix} \label{C11vb} \\
  \mathbf{C_{12}} \cdot \boldsymbol{v_j} &= \begin{pmatrix}
	b_2 + b_r \\
	b_3 \\
	b_4 \\
	\vdots \\
    b_{r-1}\\
	b_r
\end{pmatrix} \label{C12vb}
\end{align}
\end{subequations}

Considering the zero rows in $C_{21}$ and $C_{22}$, we also have
\begin{subequations}
\begin{align}
  a_1 + a_{r-1} + a_r &= 0, \label{azero}\\
  b_1 + b_{r-1} + b_r &= 0. \label{bzero}
\end{align}
\end{subequations}
According to Eqs.~(\ref{abrelations}) and (\ref{azero}), the following equation holds,
\begin{equation}\label{brelation}
  b_2 + b_{r} + b_{r-3}+b_{r-1} = 0.
\end{equation}
Next, we show that for the column pair $(C_{11}\cdot v_i, C_{12}\cdot v_j)$, no 1-initiated subsequence in one column matches the corresponding subsequence in the other. We designate this property as the backward consistency condition. First, note that the first entries of $C_{11}\cdot v_i$ and $C_{21}\cdot v_j$ both equal $b_2+b_r$. When $b_2+b_r = 1$, the column distinctness in B immediately guarantees the backward consistency condition. Further, if the second entry of $C_{11}\cdot v_i$ (or $C_{12}\cdot v_j$) equals $1$, then the column distinctness in B again directly satisfies the backward consistency condition.
Consequently, our analysis focuses on 1-initiated subsequences beginning at the third entries of $C_{11}\cdot v_i$ and $C_{21}\cdot v_j$, where we first establish a fundamental observation.

Let integers $n$, $m$, $p$ satisfy $1\le m\le p$, $n\le r-7$ and $n+6m< r$, where $r=3k+1$ and $k=2p+1$ (odd). If the subsequence alignment condition:
 \emph{the vector segment starting at row $((7-n)+6(p-m))_{th}$ row of $C_{12}\cdot v_j$ matches the corresponding subsequence in $C_{11}\cdot v_i$ position-wise,} holds, then the following equality must be satisfied:
\begin{equation}\label{Rrelation}
b_{r-n}=b_{r-n-6m},~ m\in [1, p].
\end{equation}
The relation in
Eq.~(\ref{Rrelation}) admits a recursive derivation. Without loss of generality, we examine the case where $n=0$, $m=1$ and $p=2$, the $13_{th}$ row ($=((7-0)+6(2-1))$) of $C_{12}\cdot v_j$ is $b_{r-2}$, if
\begin{equation*}
\begin{cases}
b_{r-2}=b_{r-6}+b_{r-4}\\
b_r=b_{r-4}+b_{r-2}
\end{cases}
\end{equation*}
$b_{r}=b_{r-6}$ can be obtained.
For the case $n=0$, $m=2$, the $7_{th}$ row of $C_{12}\cdot v_j$ is $b_{r-8}$, if
\begin{equation*}
\begin{cases}
b_{r-8}=b_{r-12}+b_{r-10}\\
b_{r-6}=b_{r-10}+b_{r-8}\\
b_{r-2}=b_{r-6}+b_{r-4}\\
b_r=b_{r-4}+b_{r-2}
\end{cases}
\end{equation*}
we have $b_{r}=b_{r-6}=b_{r-12}$. In fact, for $p=m$ and $n\in \{0,1,2,3\}$, we obtain the fundamental mappings:
\begin{equation}\label{odd cases}
\begin{cases}
b_r=b_4; & {\text{(Subsequence alignment initiates at row 7)}}\\
b_{r-1}=b_3; & {\text{(Row 6 initiation)}} \\
b_{r-2}=b_2; & {\text{(Row 5 initiation)}}\\
b_{r-3}=b_1; & {\text{(Row 4 initiation)}}
\end{cases}
\end{equation}

For the case where $k=2p$ (even), the analogous relation holds when subsequences align starting from row
$((4-n)+6(p-m))_{th}$ row of $C_{12}\cdot v_j$. Taking $p=m+1$, we have
\begin{equation}\label{even cases}
\begin{cases}
b_r=b_7; & \text{(Row 10 initiation)}\\
b_{r-1}=b_6; & \text{(Row 9 initiation)}\\
b_{r-2}=b_5; & \text{(Row 8 initiation)}\\
b_{r-3}=b_4; & \text{(Row 7 initiation)}
\end{cases}
\end{equation}
When $p=m=1$ (i.e. $r=7$), only $n=0$ satisfies $n+6m < r$, yielding
\begin{equation}\label{br=b1}
b_r=b_1;~\text{(Row 4 initiation)}
\end{equation}

\subsubsection{Proof of no 1-initiated subsequence starting from the $l$-th entry ($l\ge 3$) of $C_{12}v_j$ matches the corresponding subsequence in $C_{11}v_i$}
We first consider the case $k=2p+1$ (odd) for $p\ge 0$ (recall that $r=3k+1$).
It is evident that if $p=0$, $r=4$, the condition $C_{12}v_j = (001)^{T}$ (i.e. $b_2+b_4=0$, $b_3=0$, $b_4=1$) immediately imply $C_{11}v_i = (000)^{T}$, satisfying the proposition.
For cases $p\ge 1$,
\begin{enumerate}[label=(\roman*)]
  \item $l=3$: Given $b_2+b_r=0$, $b_3=0$ and $b_4=1$, Eq.~(\ref{odd cases}) yields $b_2+b_{r-2}+b_{r-1}=0$, forcing the third entry of $C_{11}\cdot v_i$ to vanish. 
  \item $l=4$: The preceding case nullifies the first three entries of $C_{11}v_i$. The full column rank of $B$ then guarantees the result.
  \item $l=5$: Combining Eq.~(\ref{odd cases}) with $b_2+b_r=0$ gives $b_2+b_4=0$, ensuring the fifth entry of $C_{11}v_i$ is zero.
  \item $l\ge 6$: For the $l$-th element of $C_{12}v_j$ being 1 with zeroed first $l-1$ entries, the $l$-th element of $C_{11}v_i$ satisfies $b_{l-1}+b_{l-3} = 0$.
\end{enumerate}

For $k=2p$ (even) with $p\ge 1$, we analyze potential subsequence matches:
\begin{enumerate}[label=(\roman*)]
  \item $l=3$: Given $b_2+b_r=0$ and $b_3=0$, assume a matching subsequence starting at the 3rd entry. Then:
\begin{align}
b_5&=b_3+b_1=b_1 \quad \text{(from recurrence)}, \nonumber\\
b_2+b_{r-2}&=b_1+b_5=0 \quad \text{(via Eq.~(\ref{even cases}))}. \nonumber
\end{align}
Thus, the first two entries of $C_{11}v_i$ must vanish. The full column rank of $B$ then precludes such matches.
  \item $l=4$: With $b_2+b_r=0$, $b_3=0$, and $b_4=0$, assume a 4th-entry match. This requires:
\begin{align}
    b_5 &= b_3 + b_1 = 1 \Rightarrow b_1=1, \nonumber\\
    b_6 &= b_2 + b_4 = b_2 \quad \text{(consistency)}. \nonumber
\end{align}
However, combining Eq.~(\ref{brelation}), $b_2+b_r=0$, and Eq.~(\ref{even cases}) yields:
\begin{equation}
    b_6 = b_4 = 0 \Rightarrow b_2 = 0, \nonumber
\end{equation}
while Eq.~(\ref{br=b1}) demands:
\begin{equation}
    b_r = b_1 = 1 \Rightarrow b_2 = 1 \quad \text{(contradiction)} \nonumber
\end{equation}

\item $l=5$:
Conditions $b_2+b_r=0$, $b_3=b_4=b_5=0$, and $b_6=1$ imply:
\begin{equation}
    b_2 = 1 \quad \text{(from $b_6=b_4+b_2$)} \nonumber
\end{equation}
Yet Eq.~(\ref{even cases}) requires:
\begin{equation}
    b_2 = b_r = b_7 = b_3 + b_5 = 0 \quad \text{(contradiction)}\nonumber
\end{equation}

\item $l\geq 6$:
The argument parallels the odd-$k$ case, with identical conclusions.
\end{enumerate}

\subsubsection{Proof of no 1-initiated subsequence starting from the $l$-th entry ($l\ge 3$) of $C_{11}v_i$ matches the corresponding subsequence in $C_{12}v_j$}
For $k=2p+1$ with $p\ge 0$, we again first consider the case $p=0$, $r=4$. $C_{11}v_i=(001)^{T}$ implies $b_2=b_4$ and $b_3=0$. However, combining with Eqs.~(\ref{bzero}) and (\ref{brelation}), we obtain $C_{12}v_j=(000)^{T}$.
For $p\ge 1$,
\begin{enumerate}[label=(\roman*)]
  \item $l=3$: Given $b_2+b_r=0$, $b_2=b_{r-2}$ and $b_{r-1}=1$, Eq.~(\ref{odd cases}) yields $b_3=0$,  nullifying the first two entries of $C_12v_j$. The full column rank of $B$ then guarantees the result.
  \item $l=4$: Eqs.~(\ref{brelation}) and (\ref{odd cases}) jointly require $b_1+b_3=0$, preventing the fourth entry of $C_{11}v_i$ from being $1$.
  \item $l=5$: The combination of Eq.~(\ref{odd cases}) with $b_2+b_r=0$ gives $b_2+b_4=0$, which excludes the possibility of the fifth entry being 1.

  \item $l\ge 6$: From Eq.~($\ref{C11vb}$), we have $b_2=b_{r-2}$ and $b_{r-1}=0$. Eq.~(\ref{odd cases}) consequently implies $b_3=b_{r-1}=0$. On the other hand, the matrix-vector product structure
\begin{equation}\label{C11vi}
\mathbf{C_{11}} \boldsymbol{v_i} = \begin{pmatrix}
	a_1 \\
	a_1+a_2 \\
	a_2+a_3 \\
	\vdots \\
    a_{r-2}+a_{r-1}
\end{pmatrix}
\end{equation}
requires that a 1-initiated subsequence starting at position $l$ satisfies $a_1=a_2=\cdots = a_{l-1}=0$ and $a_l=1$. Combined with the recurrence relation $a_i=b_{i-1}+b_{i-2}$ ($i\ge 3$) obtained from Eq.~(\ref{abrelations}) and Eq.~(\ref{C11vb}), we derive $b_1=b_2=\cdots=b_{l-2}=0$ and $b_{l-1}=1$.
Assuming the $l$-th entry of $C_{12}v_j$ equals 1 (i.e., $b_{l+1}=1$), the 6-term subsequence $(b_{l-4},...,b_{l+1})$ takes the form $(0,0,0,1,?,1)$. However, the 6-step periodicity in Eq.~(\ref{Rrelation}) prohibits the last four entries of $C_{12}v_j$ from all vanishing - contradicting the requirements from Eq.~(\ref{odd cases}):
$b_{r-3} = b_1 = 0,\ b_{r-2} = b_2 = 0,\ b_{r-1} = b_3 = 0,\ b_r = b_4 = 0$.
\end{enumerate}

For $k=2p$ (even) with $p\ge 1$, we analyze potential subsequence matches:
\begin{enumerate}[label=(\roman*)]
  \item $l=3$: Given $b_2+b_r=0$ and $b_2+b_{r-2}+b_{r-1}=1$ (3rd entry of $C_{11}v_i$), we derive $b_{r-1}=1$. Eq.~(\ref{even cases}) implies $b_1=b_r$, leading to $b_1+b_{r-1}+b_{r}=1$ which contradicts Eq.~(\ref{bzero}). 
  \item $l=4$: Assuming a matching subsequence initiating at the fourth entry, we have $b_2=b_r=b_{r-2}$, $b_{r-1}=0$ and $b_5=b_1+b_3=1$.
    From Eq.~\ref{even cases}, we obtain:
   \begin{equation}\nonumber
        b_5 = b_{r-2} = 1 \implies b_2 = b_r = b_{r-2} = 1 \label{eq:implications}
    \end{equation}
        From Eq.~(\ref{bzero}):
        \begin{equation}\nonumber
            b_1 + b_r + b_{r-1} = 0 \implies b_1 = 1 \quad (b_r=1, b_{r-1}=0)
        \end{equation}
        consequently yielding $b_3 = 0$.
From Eq.~(\ref{brelation}) with $b_{r-1} = 0$, we have
            $b_{r-3} = 0$, and
       from Eq.~(\ref{even cases}), we obtain $b_4 = 0$.
       Thereby, the vanishing first three entries in both $C_{11}v_i$ and $C_{12}v_j$ would render the columns identical, violating the full column rank property of matrix $B$.


  \item $l=5$: The conditions $b_2+b_{r-2}=0$ and $b_2+b_{r-2}+b_{r-1}=0$ imply $b_{r-1}=0$. Eq.~(\ref{even cases}) then gives $b_6=b_{r-1}=0$, preventing the 5th entry of $C_{12}v_j$ from being 1.
   \item $l\geq 6$: The proof follows the same methodology as the odd-$k$ case.
\end{enumerate}

In summary, for the Case 3: Different Matrices, Different Vector. For distinct non-zero vectors $v_i, v_j \in V_r$ with $C_{21}\cdot v_i = C_{22}\cdot v_j$, the 1-initiated subsequences in $C_{12}v_j$ and $C_{11}v_i$ cannot match.



\end{document}